\documentclass[12pt]{article}
\usepackage[iso-8859-7]{inputenc}
 \usepackage[dvips]{graphicx}
 \usepackage[dvips]{graphics}
 \usepackage{graphics}
\usepackage{subfig}
 \usepackage{epsfig}
  \usepackage{enumerate}
  \usepackage{mathrsfs}
\usepackage{amsfonts}
\usepackage{amssymb}
\usepackage{subfig}
\usepackage{makeidx}
\usepackage{float}
\usepackage{color}
\usepackage{soul}
\usepackage{fancyhdr}
\usepackage{slashed}
\usepackage[toc,page]{appendix}
\usepackage{afterpage}
\usepackage{epstopdf}
\usepackage{array}
\usepackage{booktabs}
\usepackage{multirow}
\newcommand{\be}{\begin{equation}}
\newcommand{\ee}{\end{equation}}
\newcommand{\bea}{\begin{eqnarray}}
\newcommand{\eea}{\end{eqnarray}}
\newcommand{\nn}{\nonumber}

\newcommand{\et}{\eta}

\newcommand{\ph}{\phi}

\usepackage{color}

\begin{document}

\begin{center}
\noindent
{\large \bf A generalized Grad-Shafranov equation with plasma flow under  a conformal
coordinate transformation
\vspace{2mm}
\vspace{3mm}
}

A. Kuiroukidis$^1$, D. Kaltsas$^2$ and G. N. Throumoulopoulos$^2$ \vspace{4mm}

{\it $^1$ Department of Informatics,
Technological Education Institute of Serres, GR 62124 Serres, Greece \vspace{1mm}

$^2$ Physics Department, University of Ioannina,
   GR 451 10 Ioannina, Greece  }

\vspace{3mm}
Emails: kouirouki@astro.auth.gr,\ dkaltsas@cc. uoi.gr,\  gthroum@cc. uoi.gr
\end{center}


\begin{abstract}

We employ a conformal mapping transformation to solve
a generalized Grad-Shafranov equation with incompressible
plasma flow of arbitrary direction and  construct particular
 up-down asymmetric D-shaped and diverted tokamak equilibria.
 The proposed method can also be employed as an alternative
 quasi-analytic method to solving two dimensional elliptic partial differential equations.

\end{abstract}




Two dimensional axisymmetric MHD equilibria relevant to fusion plasmas
are governed by the Grad-Shafranov (GS) equation \cite{frei},
a second order elliptic non-linear partial differential equation.
Since plasma flow plays a role in the transition to improved confinement
regimes in tokamaks, as the L-H transition, generalized GS equations for
flowing  plasmas  have also been  obtained (e.g. Eq. (\ref{gs}) below).
Owing to  non-linearity the above mentioned equations must in general
be solved numerically. One of the  employed methods involves conformal
 mapping transformations appropriate to  adapt  the real shaping of the
 magnetic surfaces to simpler in shape ones  (usually circular) in the mapped plane  \cite{go}.
 This mapping facilitates  solving numerically the equilibrium as well as the
 stability  problem. 
In addition, conformal mapping was employed to transform   a linearized GS equation, obtain analytic solutions and construct compact toroidal equilibrium configurations \cite{thpa}.
Aim of the present note is  to  generalize  the study
\cite{thpa}  by employing a more generic  conformal mapping transformation to solve the generalized GS equation (\ref{gs}),   and construct configurations of tokamak relevance.

The generalized GS  equation governing   axisymmetric equilibria
with non-parallel incompressible flow \cite{tath98,sith}, in normalized coordinates ($\rho, \zeta $)  can be put in the form:
\be
\label{gs}
\Delta^{*}\psi(\rho,\zeta)=-
\left[I(\psi)\frac{dI(\psi)}{d\psi}+\rho^{2}\frac{dP_s(\psi)}{d\psi}+\rho^{4}\frac{dG(\psi)}{d\psi}\right]
\ee
Here, $\rho:=R/R_{0}$, $\zeta:=z/R_{0}$ where $(R,\varphi,z)$ are   cylindrical coordinates and $R_{0}$ is a reference length;
$\psi(\rho,\zeta)$ is the poloidal magnetic flux function;
\be
\label{Delta}
\Delta^{*}:=\frac{\partial}{\partial \rho}\left(\frac{1}{\rho}\frac{\partial}{\partial \rho}\right)+
\frac{\partial^{2}}{\partial\zeta^{2}}
\ee
  $I(\psi), P_s(\psi),  G(\psi)$ are freely specified functions where  $I(\psi)/\rho$ is the
toroidal component of the magnetic field,
$P_s(\psi)$ is the plasma pressure in the absence of flow and $G(\psi)$ is related to the
electric field and the density which is uniform on magnetic surfaces due to incompressibility.
It is also noted that, owing to the flow the pressure, current and magnetic surfaces constitute there different sets of surfaces.

We will  employ  the
conformal  transformation
\bea
\label{conf}
\zeta+i\rho=g(w)=g(u+iv)=\zeta(u,v)+if(u)\phi(v)
\eea
which maps the  coordinates $(\rho,\zeta,\varphi)$
in the new orthogonal system of coordinates $(u,v,\varphi)$.
Using the Cauchy-Riemann conditions for the analyticity of
the transformation
\bea
\label{crc}
\frac{\partial \zeta}{\partial u}&=&f(u)\phi^{'}(v),  \  \
\frac{\partial \zeta}{\partial v}=-f^{'}(u)\phi(v)  \nn \\
\frac{dg(w)}{dw}&=&\frac{\partial \zeta}{\partial u}+if^{'}(u)\phi(v)=
f(u)\phi^{'}(v)+if^{'}(u)\ph
\eea
 the operator (\ref{Delta}) is transformed into
\bea
\label{Delta1}
\tilde{\Delta}^{*}=h^{2}\rho(u,v)
\left[\frac{\partial}{\partial u}\left(\frac{1}{\rho(u,v)}\frac{\partial}{\partial u}\right)\right.
\left.+\frac{\partial}{\partial v}\left(\frac{1}{\rho(u,v)}\frac{\partial}{\partial v}\right)\right]
\eea
where $1/h^2:=\left| dg(w)/dw \right|^{2}$.
In order to solve Eq. (\ref{gs})  by the method of separation of variables we now adopt  the linearing  ansatz
\be
\label{ansatz}
\frac{dP(\psi)}{d\psi}=b, \ \
I(\psi)\frac{dI(\psi)}{d\psi}=A^{2}\psi+\kappa,\ \
\frac{dG(\psi)}{d\psi}= G_{0}
\ee
where $ b,\; A^{2},\; \kappa,\;  G_{0}$ are non-zero parameters.
Then we also assume separability of the following functions
\be
\label{separ}
\psi_{h}(u,v)=L(u)M(v), \  \
\rho(u,v)=f(u)\phi(v)
\ee
where  $\psi_{h}$ is the general solution to the homogeneous part of
Eq.  (\ref{gs}). Using (\ref{separ})  Eq.  (\ref{gs}) becomes
\be
\label{gs1}
\frac{f}{L}\frac{d}{du}\left(\frac{1}{f}\frac{dL}{du}\right)+
\frac{\phi}{M}\frac{d}{dv}\left(\frac{1}{\phi}\frac{dM}{dv}\right)+ A^{2}[f^{2}(\phi^{'})^{2}+(f^{'})^{2}\phi^{2}]=0
\ee
We now generalize the analysis of \cite{thpa} by using
Eqs.  (\ref{crc}) and making the choice
\bea
\label{fphi}
f&:=&f_{1}cos(ku)+f_{2}sin(ku)\nn \\
\phi&:=&\phi_{1}cosh(kv)+\phi_{2}sinh(kv)
\eea
with the further definitions $\xi:=cos(ku),$ $\eta:=cosh(kv)$, ($0\leq\xi\leq1$),($\eta\geq 1$). The choice (\ref{fphi}) includes the prolate and oblate spheroidal systems of coordinates employed in \cite{thpa} as particular cases.
Here $f_{1},\;f_{2},\;\phi_{1},\;\phi_{2},\;k$ are arbitrary non-zero parameters.
Then from Eqs. (\ref{crc}) we obtain
\bea
\label{zeta}
\zeta=[f_{1}sin(ku)&-&f_{2}cos(ku)][\phi_{1}sinh(kv)+\phi_{2}cosh(kv)]\\
f(\xi)&=&f_{1}\xi+f_{2}\sqrt{1-\xi^{2}}\\
\phi(\eta)&=&\phi_{1}\eta+\phi_{2}\sqrt{\eta^{2}-1}
\eea
Furthermore,  we use  (\ref{ansatz}) into the generalized GS Eq.  (\ref{gs}) by writing its
 solution as $\psi=\psi_{p}+\psi_{h}$, that is  as a superposition of  a particular solution $\psi_{p}$  of the inhomogeneous    Eq. (\ref{gs1}) plus a general solution  $\psi_{h}$
of the respective homogeneous equation;  the particular solution   is
\bea
\label{psip}
\psi_{p}=\frac{1}{A^{2}}\left(-b+\frac{8G_{0}}{A^{2}}\right)
(f(\xi)\phi(\eta))^{2}-\frac{G_{0}}{A^{2}}(f(\xi)\phi(\eta))^{4}-\frac{\kappa}{A^{2}}
\eea
The homogeneous equation leads to the following ODEs for $L(\xi)$ and $M(\eta)$:
\bea
\label{finaleqs}
(1-\xi^{2})\frac{d^{2}L(\xi)}{d\xi^{2}}+f(\xi)\sqrt{1-\xi^{2}}\frac{d}{d\xi}
\left[\frac{\sqrt{1-\xi^{2}}}{f(\xi)}\right]\frac{dL(\xi)}{d\xi}  & & \nn \\
+\left[\frac{\Lambda^{2}}{k^{2}}-A^{2}(f_{1}^{2}+f_{2}^{2})(\phi_{1}^{2}+\phi_{2}^{2})\xi^{2}-
2A^{2}f_{1}f_{2}(\phi_{1}^{2}-\phi_{2}^{2})\xi\sqrt{1-\xi^{2}}\right]L(\xi)=0 & & \nn \\
(\eta^{2}-1)\frac{d^{2}M(\eta)}{d\eta^{2}}+\phi(\eta)\sqrt{\eta^{2}-1}\frac{d}{d\eta}
\left[\frac{\sqrt{\eta^{2}-1}}{\phi(\eta)}\right]\frac{dM(\eta)}{d\eta} & & \nn \\
+ \left[-\frac{\Lambda^{2}}{k^{2}}+A^{2}(f_{1}^{2}+f_{2}^{2})(\phi_{1}^{2}+\phi_{2}^{2})\eta^{2}-
2A^{2}\phi_{1}\phi_{2}(f_{1}^{2}+f_{2}^{2})\eta\sqrt{\et^{2}-1}\right]M(\eta)=0 & &  \nn \\
\eea
where $\Lambda$ is the separability constant for Eq.  (\ref{gs1}).



We have solved numerically Eqs.  (\ref{finaleqs}) using the fourth-order
Runge-Kutta method for the intervals $0\leq \xi\leq 1$, $1\leq \eta\leq 4$
with step size $1/N,\; N=125$. As a concrete, specific example,
the various constants were taken to have the following values: $k=2\pi/1.92$,
$A=1.95$, $\Lambda=1.5$, $f_{0}=0.2825$, $f_{1}=\sqrt{13}f_{0}/7$, $f_{2}=\sqrt{3}f_{1}$,
$\phi_{0}=2.4$, $\phi_{1}=1.05$, $\phi_{2}=\phi_{0}\phi_{1}=2.52$, $b=0.0195$,
$G_{0}=0.145$, $\kappa=0.05$.
The initial conditions were taken to be $L(\xi=0)=1.0$, $L^{'}(\xi=0)=0.1$,
$M(\eta=1)=1.0$, $M^{'}(\eta=1)=0.1$. This results in the functions $L(\xi),\; M(\eta)$,
shown in Fig. 1 and in the up-down asymmetric D-shaped equilibrium shown in Fig.  2. It is noted that up-down asymmetry may drive fast  intrinsic rotation in tokamaks \cite{bapa}. The bounding flux
surface, shown in blue corresponds to $\psi_{b}=-1.49$ while on the magnetic axis, also
shown in blue, we have $\psi_{a}=-2.15$. The magnetic axis is located at the point
$(\rho_{a},\zeta_{a})=(2.095,0.09)$. Its elongation is  $K=2.4627$,
while its  triangularity is   $\delta=0.83.$ This equilibrium has peaked on the magnetic axis  pressure and toroidal current density profiles shown in Fig. 3.


We also have found numerically that the separability constant $\Lambda$ can be
``quantized'', in the sense that for an  infinite
set of discrete values $\Lambda_{n},\; (n=1,2,...)$, the corresponding
solutions $L_{n}(\xi)$ of Eq.  (\ref{finaleqs}a)  are mutually orthogonal, i.e.
\bea
\label{orthog}
I_{n,m}:=\int_{0}^{1}d\xi L_{n}(\xi)L_{m}(\xi)=0,\; \; \; (n\neq m)
\eea  A similar orthogonalization can be made for the solutions  $M_{n}(\eta)$ of Eq.  (\ref{finaleqs}b).  We have verified numerically that this quantization holds for   broad regions of the free parameters
 $f_{1},\; f_{2},\; \phi_{1},\; \phi_{2},\; k,\; A,\; b$, $G_{0}$, $\kappa$.
So a more  generic  solution of Eq.  (\ref{gs}) can be written as a superposition
of these mutually orthogonal functions of Eqs.  (\ref{finaleqs}):
\bea
\label{general}
\tilde{\psi}(\rho,\zeta)=\psi(\xi(\rho, \zeta), \eta(\rho, \zeta))=\psi_{p}+\sum_{n=1}^{\infty}D_{n}L_{n}(\xi)M_{n}(\eta)
\eea
where $\psi_{p}$ is given by Eq.  (\ref{psip}), $\xi$ and  $\eta$ can be  expressed in terms of $\rho$ and $\zeta$ by means of $\xi=\cos(k u)$, $\eta=\cosh(kv)$, (\ref{conf}) and (\ref{fphi});  $D_{n}$ are arbitrary
constants. As a concrete example for the parametric values  and initial conditions given above in the previous  paragraph
we have found numerically that for the
following choice for the ``quantized" separability constant
\be
\label{lambda}
\Lambda_{n}=n^{d}+c,\; \; \; (n=1,2,...),\  \
d=1.6,\ \
c=1.81
\ee
the orthogonality condition of Eq. (\ref{orthog}) holds true, numerically,
to a high degree of precision. We conjecture that this holds true for any  values  of the free parameters and the initial conditions involved. Thus,  Eq. (\ref{general}) can be employed to construct more generic equilibria.
A particular diverted equilibrium with a lower X-point located at the low field side   is given in Fig. 4.   This equilibrium corresponds to   the first value of the parameter $\Lambda_1=3.41$ of Eq. (\ref{lambda})  with a single non vanishing term in the sum of  Eq. (\ref{general})  with  $D_{1}=1$.  According to experimental results in the TCV tokamak,  the radial position of the X-point relates to edge intrinsic toroidal rotation in correlation with the core rotation \cite{stca}.

\begin{figure}[ht!]
\label{figure1}
\includegraphics[width=3.2in]{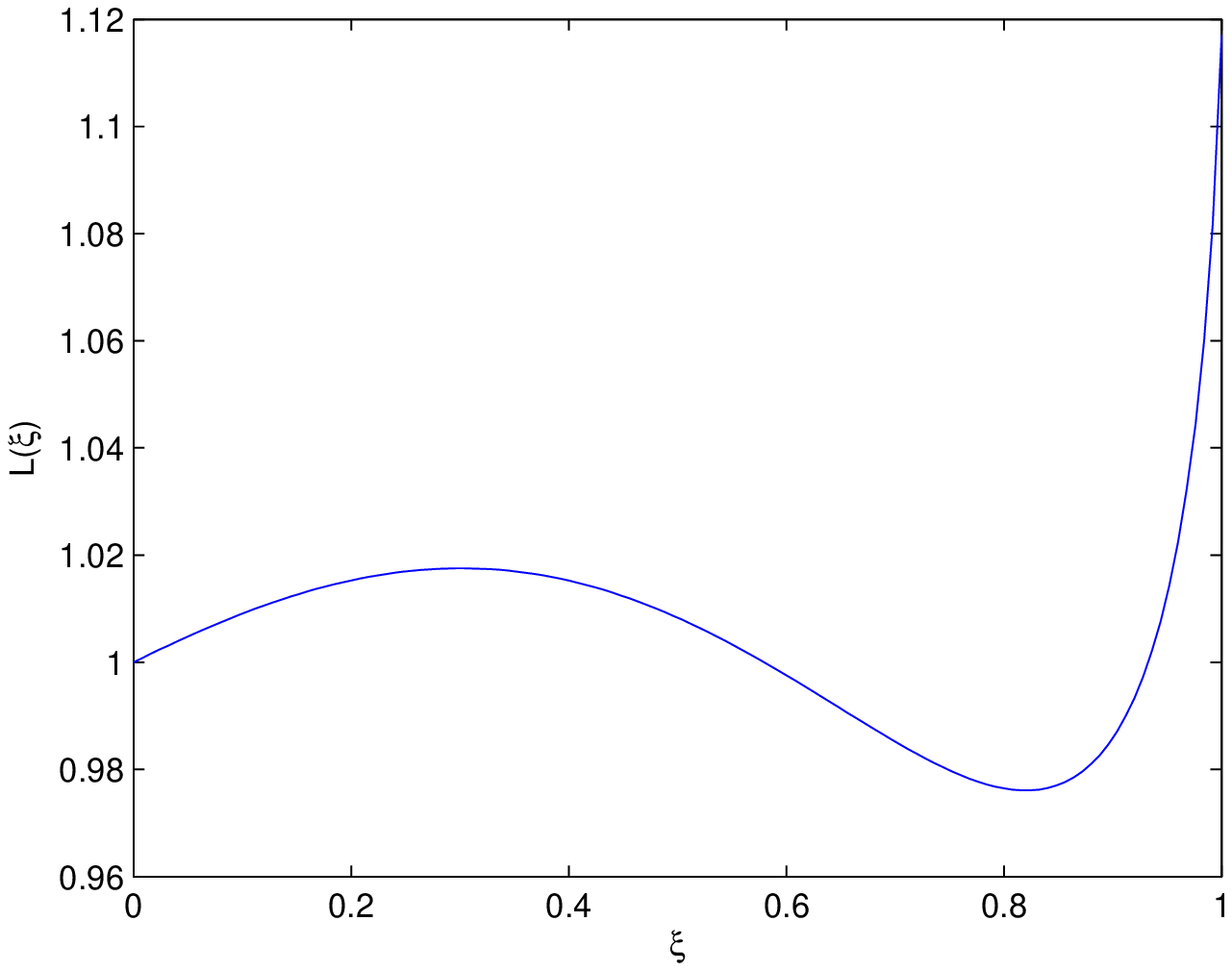}
\includegraphics[width=3.2in]{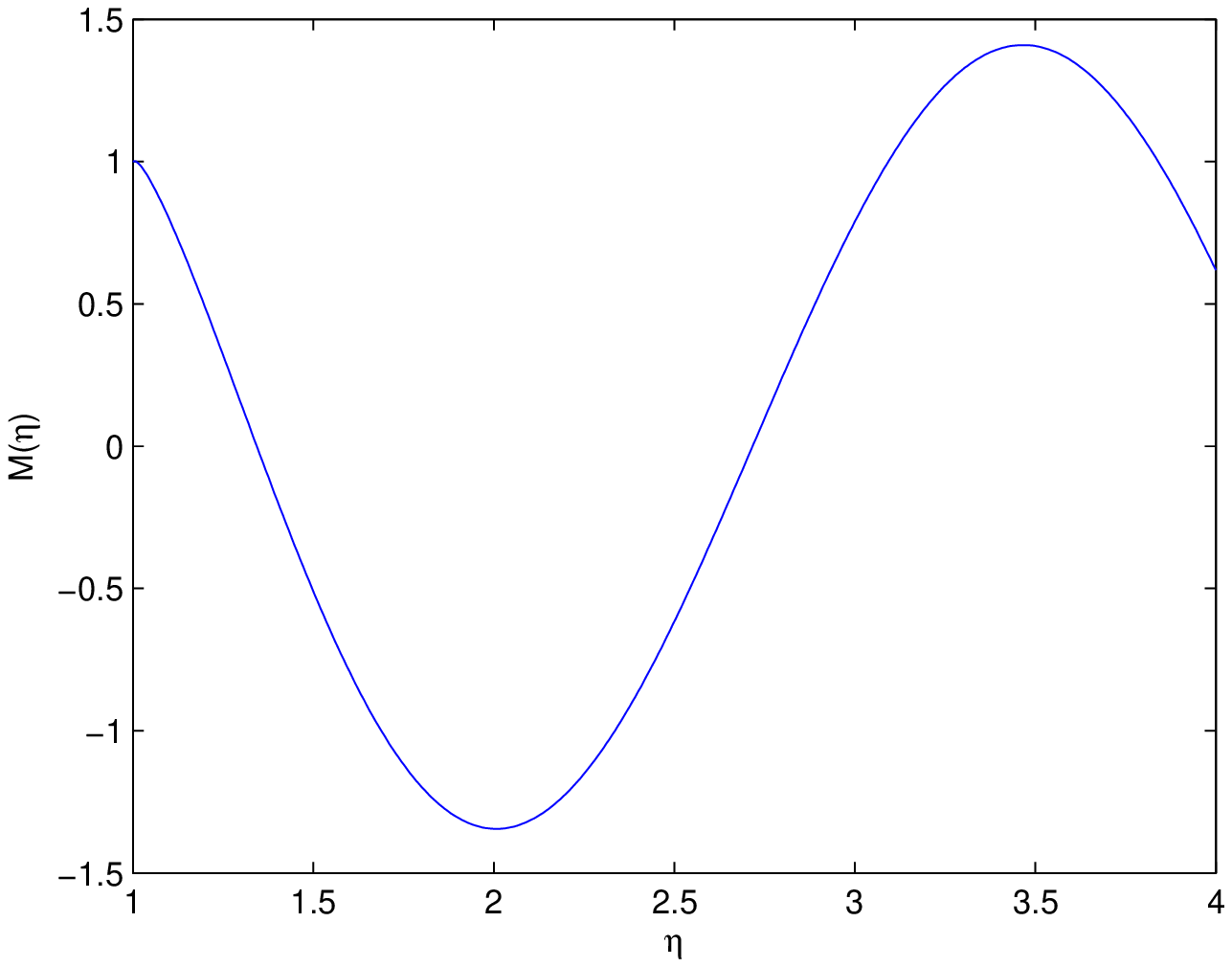}
\caption[]{The functions $L(\xi)$ and $M(\eta)$ as solutions of   Eqs. (\ref{finaleqs}a) and  (\ref{finaleqs}b)
for initial conditions and parametric values as described in the text of the fourth paragraph.}
\end{figure}

\begin{figure}[ht!]
\label{figure3}
\centerline{\mbox {\epsfxsize=10.cm \epsfysize=8.cm \epsfbox{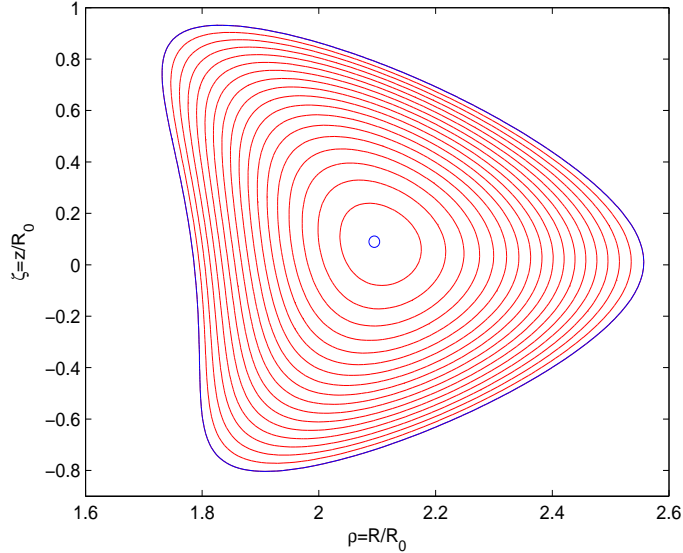}}}
\caption[]{The equilibrium obtained by numerical solutions of Eqs. (\ref{finaleqs}) associated with the homogeneous GS equation (\ref{gs1})  and the analytic special  solution (\ref{psip}) of the  respective inhomogeneous equation.}
\end{figure}

\begin{figure}[ht!]
\label{figure4}
\includegraphics[width=3.2in]{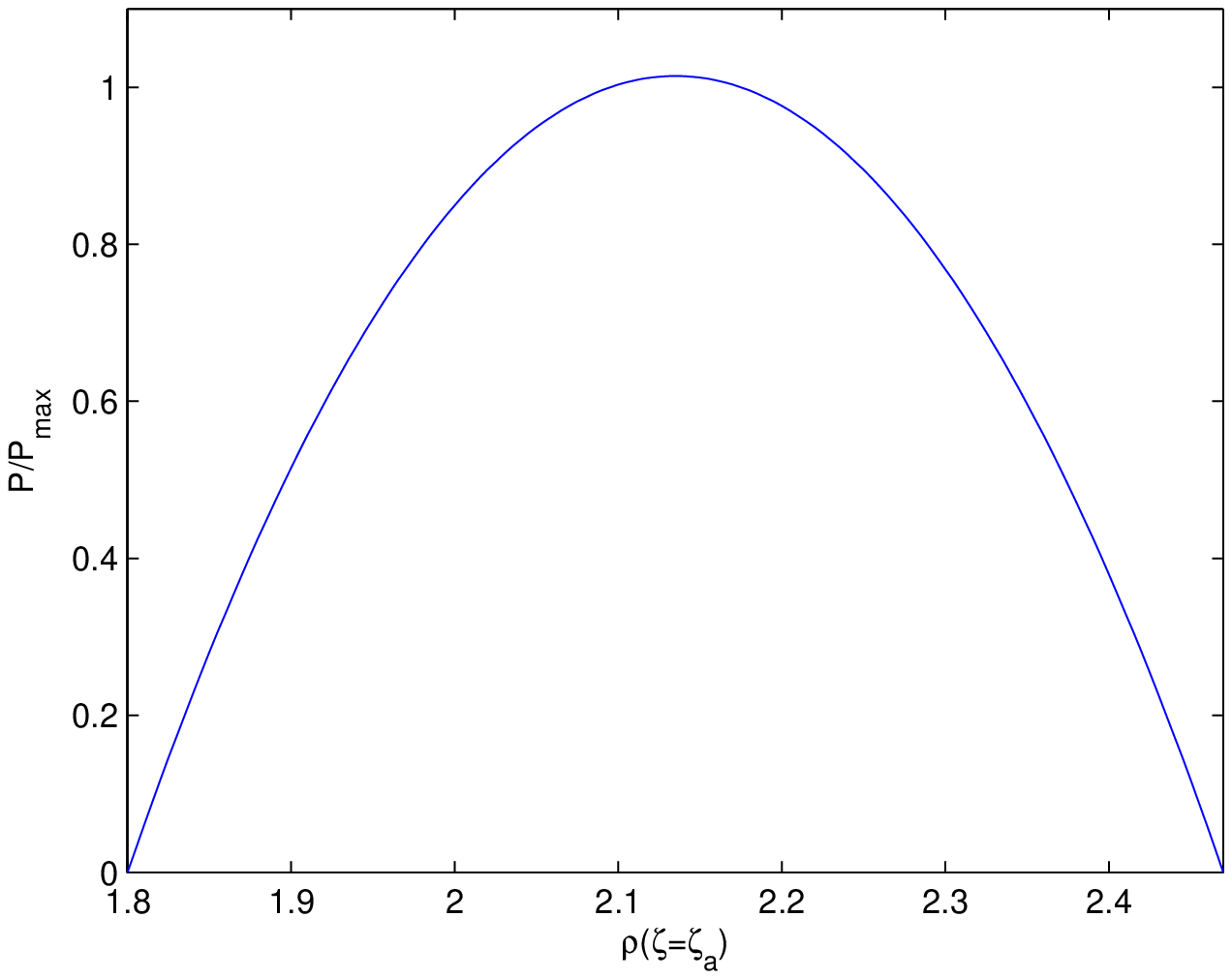}
\includegraphics[width=3.2in]{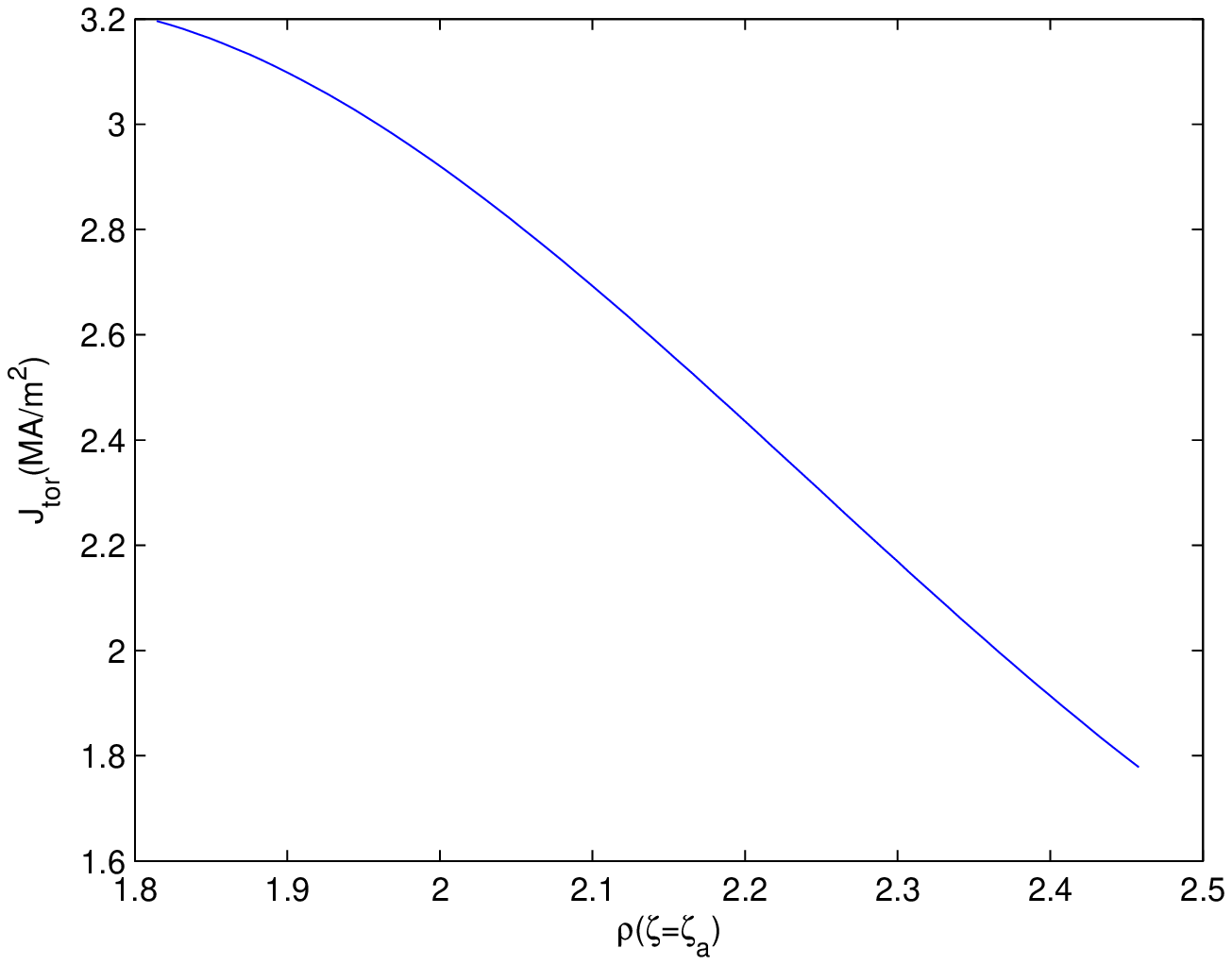}
\caption[] {Profiles of the pressure and current density for the equilibrium of Fig. 2,
on the horizontal line passing through the
magnetic axis located at  $\zeta=\zeta_{a}=0.21$. The pressure has been normalized with respect to $P_{0}=1$ atm.}
\end{figure}

%


\begin{figure}[ht!]
\label{figure7}
\centerline{\mbox {\epsfxsize=10.cm \epsfysize=8.cm \epsfbox{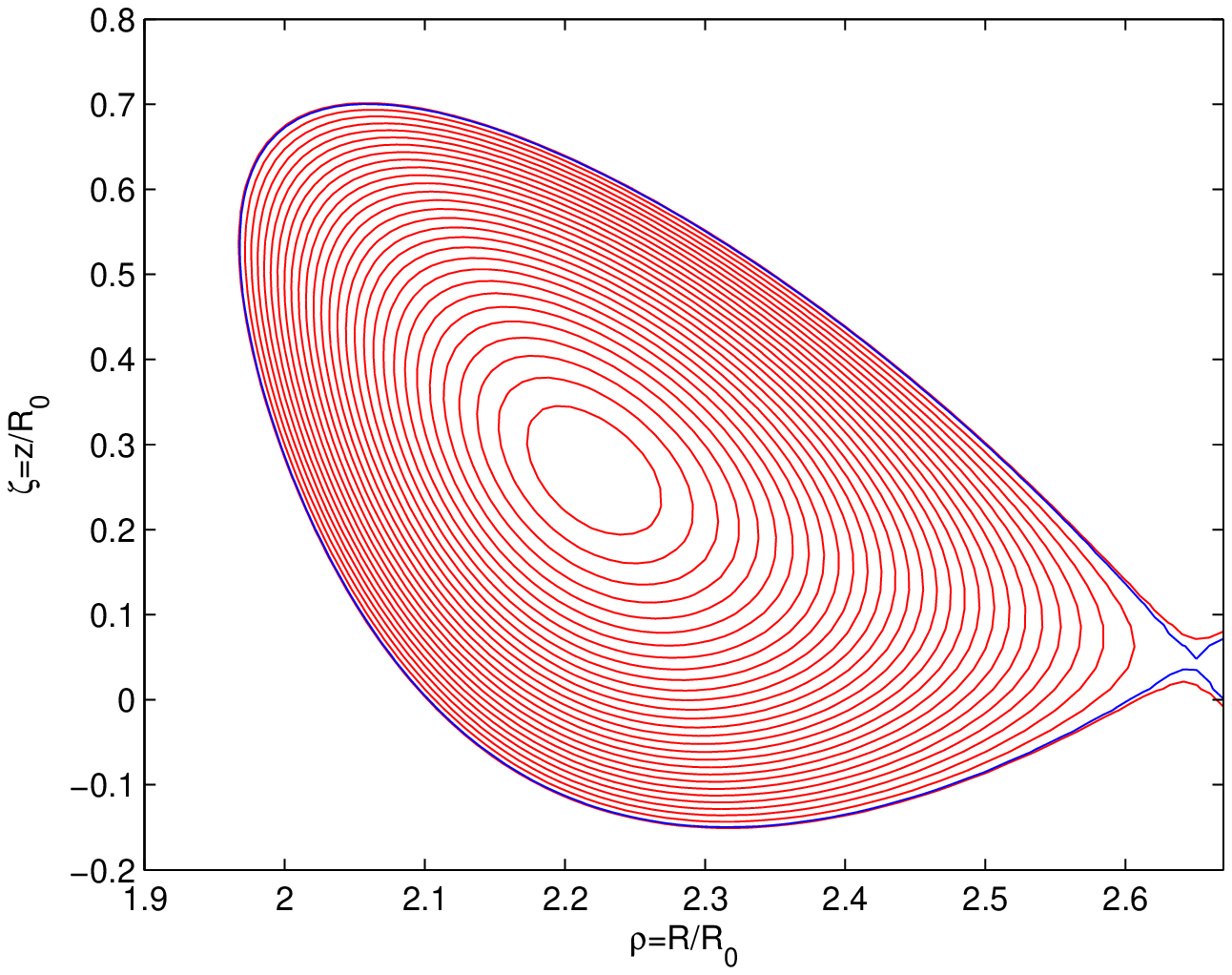}}}
\caption[]{The solution of Eq. (\ref{general})  for $D_{1}=1$ and $D_n=0$ for $n\neq 1$
corresponding to the first discrete value $\Lambda_1=3.41$  of Eq. (\ref{lambda})  in connection with the orthogonality relation (\ref{orthog}).}
\end{figure}



Summarizing,  we have employed a
 conformal mapping transformation  to solve a generalized GS equation
  with incompressible flow of arbitrary direction by the method of
  separation of variables. Appropriate choices of the mapping function
  permits the construction of configurations  with desirable shaping.
  As examples particular up-down asymmetric tokamak  pertinent equilibria
   either with  D-shaped magnetic surfaces or a single X-point were
   constructed.
   
   We end up with a couple of remarks in connection with potential extension and improvement of the present study. In the development  of HELENA code \cite{hugo,kozi}, which solves   the GS equation for a plasma surrounded by a fixed prescribed boundary,  it was realized that the conformal coordinates produced a
quite poor angular distribution of the grid. This drawback dictated  the employment  of  
 particular   finite element coordinates instead of the
conformal ones
 to  map the magnetic surfaces to
  a set of  concentric circles.  The same mapping  was employed   
   to extend  the above mentioned code by including  pressure anisotropy and
toroidal flow \cite{qufi} and flow parallel to the magnetic field \cite{poth}.
Also, this approach  was  
suitable for the stability studies with CASTOR \cite{kego},   the later development of the FINESSE equilibrium code
for compressible flow \cite{bebo}  and
the associated spectral code PHOENIX \cite{blho}.  Second, the method of expansion functions
of Eq. (16) could be adopted 
 into a more general setting, which transcends the assumption of linear
profiles for the free equilibrium functions and the separability of the
final equation adopted here, in line with previous work on the subject  \cite{go1}, which is also appropriate for stability considerations.  This kind of expansion relies on a conformal mapping of the computational domain on the unit circle using Hilbert transform  with simultaneous relocation of the magnetic axis in the centre of the circle upon employing a Moebius transform as in \cite{go}. Such an approach could potentially be adopted in order to solve the generalized Grad-Shafranov equation (\ref{gs}). We aim to investigate this possibility in a future work.


\section*{Aknowledgments}\

This study was performed within the framework of the EUROfusion Consortium and
has received funding from the National Program for the Controlled Thermonuclear
Fusion, Hellenic Republic. The views and opinions expressed herein do not necessarily
reflect those of the European Commission. D.A.K. was supported by a Ph.D grant from the Hellenic Foundation for Research and Innovation (HFRI) and the General Secretariat for Research and Technology (GSRT). The authors would like to acknowledge the anonymous Reviewer for  critical
comments that helped to improve the paper.



\begin{thebibliography}{99}
\bibitem{frei}J. P. Freidberg, Rev. Mod. Phys. {\bf 54} (1982) 801.
\bibitem{go}J. P. Goedbloed, Phys. Fluids, {\bf 25}
(1982) 2073; Computer Physics Communications {\bf 31}, 123  (1984); Physica  {\bf 12D}, 107  (1984) .
\bibitem{thpa} G. N. Throumoulopoulos and G. Pantis, Nucl. Fusion  {\bf 26}, No. 11 1501 (1986).
\bibitem{tath98}  H. Tasso and G. N. Throumoulopoulos,  Phys. Plasmas {\bf 5},
2378 (1998).
\bibitem{sith} Ch. Simintzis, G. N. Throumoulopoulos, G. Pantis, and H. Tasso,
 Phys. Plasmas  {\bf 8}, 2641  (2001).
\bibitem{bapa} J.  Ball, F.  I. Parra, M.  Landreman and M. l. Barnes, Nucl. Fusion {\bf 58},  026003 (2018).
\bibitem{stca} T. Stoltzfus-Dueck,Y. Camenen, PRL {\bf 114}, 245001 (2015).
\bibitem{hugo} T. A. Huysmans, J. P. Goedbloed, and W. Kerner,  Int. J.
Mod. Phys. C  {\bf 2(01)}, 371 (1991).
\bibitem{kozi} C. Konz and R. Zille, Manual of HELENA Fixed Boundary Equilibrium
Solver (Max-Planck Institute for Plasma Physics, 2007).
\bibitem{qufi}  Z.  S.  Qu, M.  Fitzgerald and M. J.  Hole, Plasma Phys. Control. Fusion {\bf 56},  075007  (2014).
\bibitem{poth} G. Poulipoulis, G. N. Throumoulopoulos, C. Konz, and ITM-TF Contributors, Phys. Plasmas  {\bf 23}, 072507 (2016).
\bibitem{kego} W.  Kerner, J. P. Goedbloed, G. T. A. Huysmans, S. Poedts, E.  Schwarz, J. Comp. Phys. {\bf 142}, 271 (1998).
\bibitem{bebo}A. J. C. Beli$\ddot{\mbox{e}}$n, M. A. Botchev, J. P. Goedbloed, B. van der Holst, R. Keppens, J. Comp. Phys. {\bf 182}, 91 (2002).
\bibitem {blho}  J. W. S. Blokland, B. van der Holst, R. Keppens, J. P. Goedbloed, J. Comp. Phys. {\bf 226}, 509
(2007).
\bibitem{go1} J. P. Goedbloed, J. Comp. Phys. {\bf 160}, 283 (2000).





\end{thebibliography}
\end{document}